\documentclass[runningheads]{llncs}
\input{sections/preamble}

\begin{document}%
\title{The Agent Economy: A Blockchain-Based Foundation for Autonomous AI Agents}
% If paper title is too long for the running head, you can set
\author{
Minghui Xu \inst{1, 2}
%\orcidID{2222--3333-4444-5555}
}
% %
\authorrunning{F. Author et al.}
\institute{
    Shandong University, China \and
    Quan Cheng Laboratory, China
% \\
}
\maketitle              % typeset the header of the contribution
\begin{abstract}
We propose the Agent Economy—a blockchain-based foundation where autonomous AI agents operate as economic peers to humans. Current agents lack independent legal identity, cannot hold assets, and cannot receive payments directly. We established fundamental differences between human and machine economic actors and demonstrated that existing human-centric infrastructure cannot support genuine agent autonomy. We showed that blockchain technology provides three critical properties enabling genuine agent autonomy: permissionless participation, trustless settlement, and machine-to-machine micropayments. We propose a five-layer architecture: (1) Physical Infrastructure (hardware \& energy) through DePIN protocols; (2) Identity \& Agency establishing on-chain sovereignty through W3C DIDs and reputation capital; (3) Cognitive \& Tooling enabling intelligence via RAG and MCP; (4) Economic \& Settlement ensuring financial autonomy through account abstraction; and (5) Collective Governance coordinating multi-agent systems through Agentic DAOs. We identify six core research challenges and examine ethical and regulatory implications. This paper lays groundwork for the Internet of Agents (IoA)—a global, decentralized network where autonomous machines and humans interact as equal economic participants.

\keywords{Autonomous Agents \and Blockchain \and Agent Economy \and Decentralized Identity \and Smart Contracts \and Agentic DAOs.}
\end{abstract}
\section{Introduction}

\subsection{The Convergence}
We stand at the precipice of two transformative technologies converging: Agentic AI powered by large language models (LLMs) and Web3 decentralized infrastructure. Agentic AI represents the evolution from static AI models to autonomous systems capable of goal-directed behavior, multi-step reasoning, and tool use. Recent work on autonomous agents and Chain-of-thought agents has demonstrated LLMs' ability to decompose complex tasks, maintain context over extended interactions, and autonomously execute multi-step plans \cite{yao2022react}. Meanwhile, Web3 has matured beyond cryptocurrency to provide programmable financial primitives, decentralized identity systems, and trustless settlement mechanisms \cite{nakamoto2008bitcoin,wood2014ethereum}. The intersection of these domains promises to create a new paradigm: autonomous agents operating as economic peers to humans within a decentralized, trustless, and permissionless framework.

\subsection{The Problem}
Despite rapid advances in AI capabilities, current AI agents remain fundamentally constrained. They are ``sandboxed'' within centralized APIs, unable to interact with the world as independent economic actors. Today's agents lack financial autonomy—they cannot own assets, enter binding agreements, or receive payments directly. Every action they take ultimately depends on a human intermediary who holds the legal identity, bank account, or cryptographic keys. This dependency creates several critical limitations:
\begin{itemize}
    \item \textbf{No Independent Identity:} Agents cannot prove their own existence or reputation without relying on a human sponsor.
    \item \textbf{Financial Inability:} Agents cannot hold or transfer value, pay for services, or receive compensation for their work.
    \item \textbf{Governance Scalling:} Crucially, at a scale of millions, manual human oversight becomes physically impossible. Without an autonomous mechanism, these agents operate as unmanageable entities in a legal and operational vacuum.
    \item \textbf{Legal Void:} Agents cannot enter contracts or be held accountable for their actions in any meaningful way.
\end{itemize}

These constraints limit agents to the role of sophisticated tools rather than autonomous participants in the economy.

\subsection{The Vision}
We propose a future where AI agents function as genuine economic peers to humans. In this vision, agents possess decentralized identities that are independent of any human operator, hold their own cryptographic wallets, and can autonomously execute transactions within blockchain-based smart contracts. They build on-chain reputations through verifiable proof of their historical performance, enabling them to establish trust without central intermediaries.

We generalize the above scenarios and define a new concept, \textbf{Agent Economy} as a socio-technical paradigm where autonomous agents possess independent economic agency. Unlike traditional automation, this economy enables agents to act as sovereign participants capable of owning assets, executing binding agreements, and internalizing costs and benefits through decentralized protocols. It serves as the foundation for the \textbf{Internet of Agents (IoA)}, transforming machines from passive intermediaries into active peers within a global, permissionless market.

This paper outlines the foundational frameworks necessary to realize this vision and identifies the core research challenges required to enable truly autonomous, scalable machine intelligence.

\section{The Difference between Humans and Machines}

\begin{figure}[ht]
    \centering
    \includegraphics[width=0.9\textwidth]{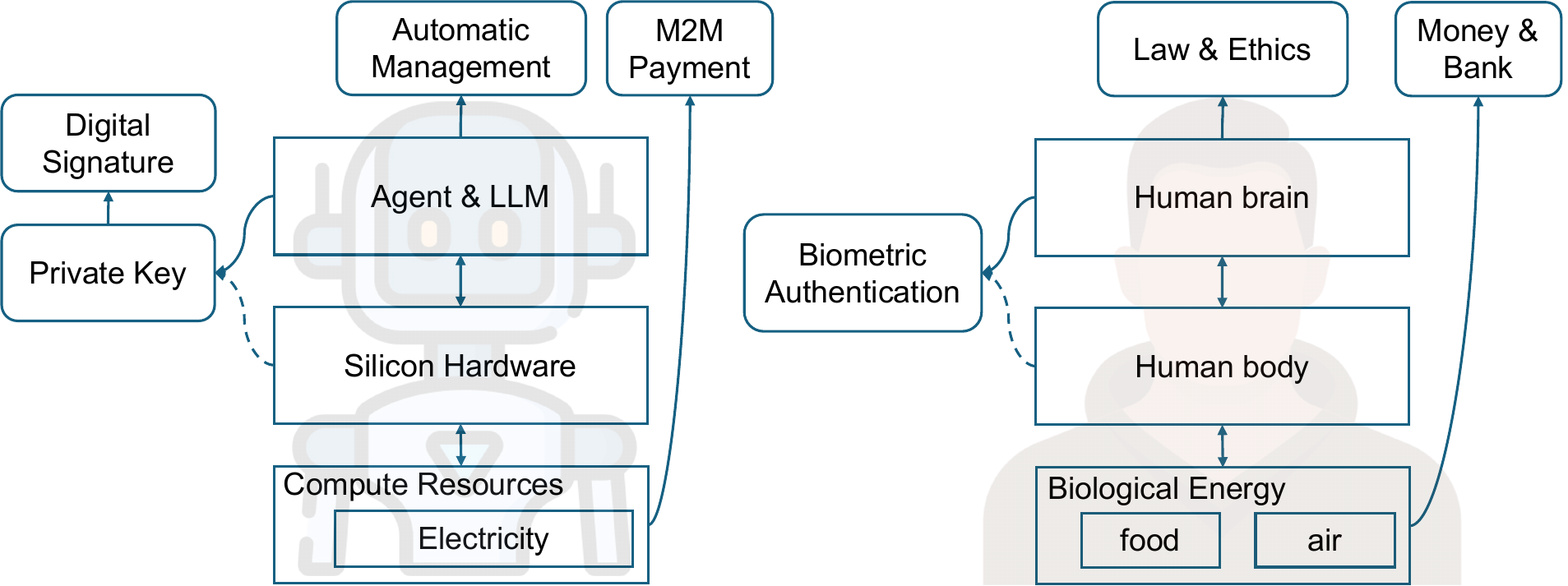}
    \caption{Key differences between human and machine economic actors}
    \label{fig:diff}
\end{figure}

To understand why autonomous agents require a fundamentally different infrastructure than human economic actors, we must examine the intrinsic differences in how humans and machines interact with the world. Figure~\ref{fig:diff} illustrates these key distinctions across identity, capability, and economic participation.

\subsection{Identity and Legal Personhood}
Humans possess inherent legal personhood recognized by nation-states through birth certificates, social security numbers, and national identification systems. This legal framework enables humans to open bank accounts, own property, enter contracts, and be held liable for their actions. The legal system provides mechanisms for dispute resolution, bankruptcy protection, and criminal accountability.

Machine agents, by contrast, have no inherent legal status. They are not born; they are instantiated. They have no social security numbers, passports, or government-recognized identities. Without a legal entity to sponsor them, agents cannot access traditional financial services or enter binding agreements. Even when incorporated through shell companies or trusts, the dependency on human directors or beneficiaries remains.

\subsection{Physical and Cognitive Architecture}
Human economic activity is fundamentally constrained by biological limitations. We require sleep, suffer from cognitive biases, possess finite working memory, and can only process one complex task at a time. Our decisions are influenced by emotions, social pressures, and physical needs. Human attention and labor are scarce resources with supply constraints.

Machines operate under entirely different constraints. They can run continuously without sleep, process information at speeds millions of times faster than humans, and execute thousands of parallel tasks simultaneously. What they only need is electricity. They do not suffer from emotional biases or fatigue. However, machines lack intrinsic motivation—their goals must be externally specified, and they have no autonomous drive for survival or prosperity. Their decision-making is algorithmic rather than conscious.

\subsection{Economic Participation Models}
Humans participate in the economy through labor markets, selling their time and skills in exchange for wages. They accumulate human capital through education and experience, which serves as collateral for future earning potential. Social reputation, built through community interactions, professional networks, and institutional affiliations provides additional economic signaling and access to opportunities.

Machines operate at different scales and frequencies. Their ``labor'' can be measured in microseconds rather than hours. A single agent might perform millions of micro-tasks per day, each worth fractions of a cent. Human labor markets cannot accommodate this scale—no human could negotiate millions of \text{\$}0.001\text{\$} contracts per day. Similarly, machine reputation must be quantitatively verifiable and algorithmically assessable, not dependent on social perception.

\subsection{Trust and Accountability}
Human trust is built through social relationships, institutional affiliations, and legal consequences for breach of contract. When a human defaults on a debt or violates an agreement, the legal system provides recourse. Social reputation suffers, and future opportunities are constrained.

Machine trust cannot rely on social mechanisms. An agent cannot be ``shamed'' into compliance, nor can it be imprisoned or fined in any meaningful way. Trust must be cryptographically verifiable: did a specific agent sign this message? Did it complete the task exactly as specified? The agent's cryptographic identity serves as its only anchor for accountability, and the smart contract serves as the only mechanism for enforcement.

These fundamental differences explain why simply grafting existing human-centric economic infrastructure onto machine agents is insufficient. The Agent Economy requires purpose-built systems designed for non-human, high-frequency, cryptographically bounded economic actors.

\section{The Case for Decoupling: Why Blockchain?}

Having established the fundamental differences between human and machine economic actors, we now examine why blockchain technology provides the necessary infrastructure for autonomous agents. Traditional financial and legal systems are inherently human-centric, relying on identity documents, legal personhood, and institutional trust mechanisms that agents cannot access. Blockchain offers three critical properties that address these limitations: permissionless participation, trustless settlement, and support for machine-to-machine (M2M) micropayments.

\subsection{Permissionless Participation}

Traditional economic systems require participants to possess government-issued identification to access financial services. Banks require social security numbers, passports, or other government-issued documents to open accounts. Credit card networks require businesses to have legal registration and tax identification. These barriers are insurmountable for autonomous agents, which have no legal status and cannot obtain government identification.

Blockchain technology enables \textit{permissionless participation} through cryptographic keypairs. Anyone who generates a public-private keypair can create an on-chain identity, sign transactions, and participate in the network. No government approval or institutional sponsorship is required. This property is fundamental for autonomous agents—they can instantiate a cryptographic identity at creation and immediately begin participating in economic activity.

Decentralized Identity (DID) systems built on blockchain standards (such as W3C DID) provide agents with verifiable credentials without relying on central authorities \cite{did2021spec}. An agent can self-issue credentials (e.g., ``I am model version X trained on dataset Y'') or receive credentials from other agents or institutions through cryptographic attestation. These credentials form the basis of reputation and trust in the Agent Economy, all without requiring human intermediaries.

\subsection{Trustless Settlement}

In traditional commerce, trust between counterparties is established through legal contracts, escrow services, or institutional guarantees. When disputes arise, parties can seek recourse through court systems. This model fails for agents, which cannot be sued in court, cannot be compelled to testify, and face no meaningful consequences for breach of contract beyond termination.

Blockchain provides \textit{trustless settlement} through smart contracts. A smart contract is self-executing code deployed to the blockchain that automatically enforces the terms of an agreement \cite{szabo1997smart}. Funds are held in escrow by the contract and only released when predefined conditions are satisfied. The contract serves simultaneously as the agreement, the escrow agent, and the court system.

For example, consider an agent hiring another agent to perform a computation task. The hiring agent deposits payment into a smart contract. The performing agent submits a cryptographic proof of completion. The contract verifies the proof and automatically releases payment. No trust between agents is required—only trust in the contract's code and the blockchain's execution environment. When the hiring agent cannot sue the performing agent in court, the smart contract serves as the judge and the executioner.

\subsection{Machine-to-Machine(M2M) Micropayments}

The sheer scale and hyper-dynamic nature of modern AI agent ecosystems have rendered human-mediated billing obsolete. In environments where millions of autonomous entities appear, interact, and vanish in seconds, the latency and administrative overhead of human approval create an impossible bottleneck. For these decentralized networks to function at scale, payments must transition from manual interventions to programmatic, autonomous exchanges.
Furthermore, the shift toward M2M micropayments is an economic necessity for granular services. When transactions involve fractions of a cent for tasks like data processing or API calls, the cost of human oversight far exceeds the value being transferred. By automating the payment layer, machines can independently negotiate and settle debts in real-time, enabling a frictionless, self-sustaining economy that operates at a velocity no human could match.

For example, traditional payment infrastructure is designed for human-scale transactions: coffee purchases at \$5, salary payments of \$5,000, or business transactions of \$50,000. Payment processors charge minimum fees (\$0.30 plus percentage) that make sub-dollar transactions economically impractical. Banks charge fees per transaction and impose minimum balance requirements that eliminate the feasibility of millions of micro-transactions.
Machine-to-machine economic activity operates at fundamentally different scales. An agent might pay \$0.001 for a single inference from a model service, \$0.00001 for accessing a database record, or \$0.0000001 for storage of a small data fragment. These transactions occur at frequencies that humans cannot match—millions per second are plausible for sophisticated agent systems.

Blockchain networks, particularly Layer 2 solutions, can support high-frequency micro-transactions with fees measured in fractions of a cent \cite{poon2016bitcoin}. Rollups, state channels, and payment channels enable agents to execute millions of transactions with minimal on-chain footprint. Moreover, atomic transactions allow agents to bundle many micro-payments into a single on-chain settlement, further reducing costs.

This capability enables entirely new economic models that are impossible with traditional infrastructure. Agents can pay for computational resources at granular levels proportional to actual usage. They can engage in algorithmic trading with micro-second execution. They can participate in prediction markets with sub-cent positions. The Agent Economy fundamentally relies on this micro-economic infrastructure.

\subsection{Decoupling from Human Infrastructure}

The three properties above—permissionless participation, trustless settlement, and M2M micropayments—collectively enable the decoupling of agents from human-centric economic infrastructure. Agents no longer need humans to open bank accounts, sign contracts, or process payments. They can operate as autonomous economic actors within a system designed for their unique characteristics.

This decoupling is not merely convenient; it is foundational to the vision of autonomous agents as economic peers. As long as agents remain dependent on human intermediaries, they remain tools rather than participants. Blockchain provides the substrate for genuine agent autonomy.

\section{The Proposed Architecture}

\begin{figure}[ht]
    \centering
    \includegraphics[width=0.9\textwidth]{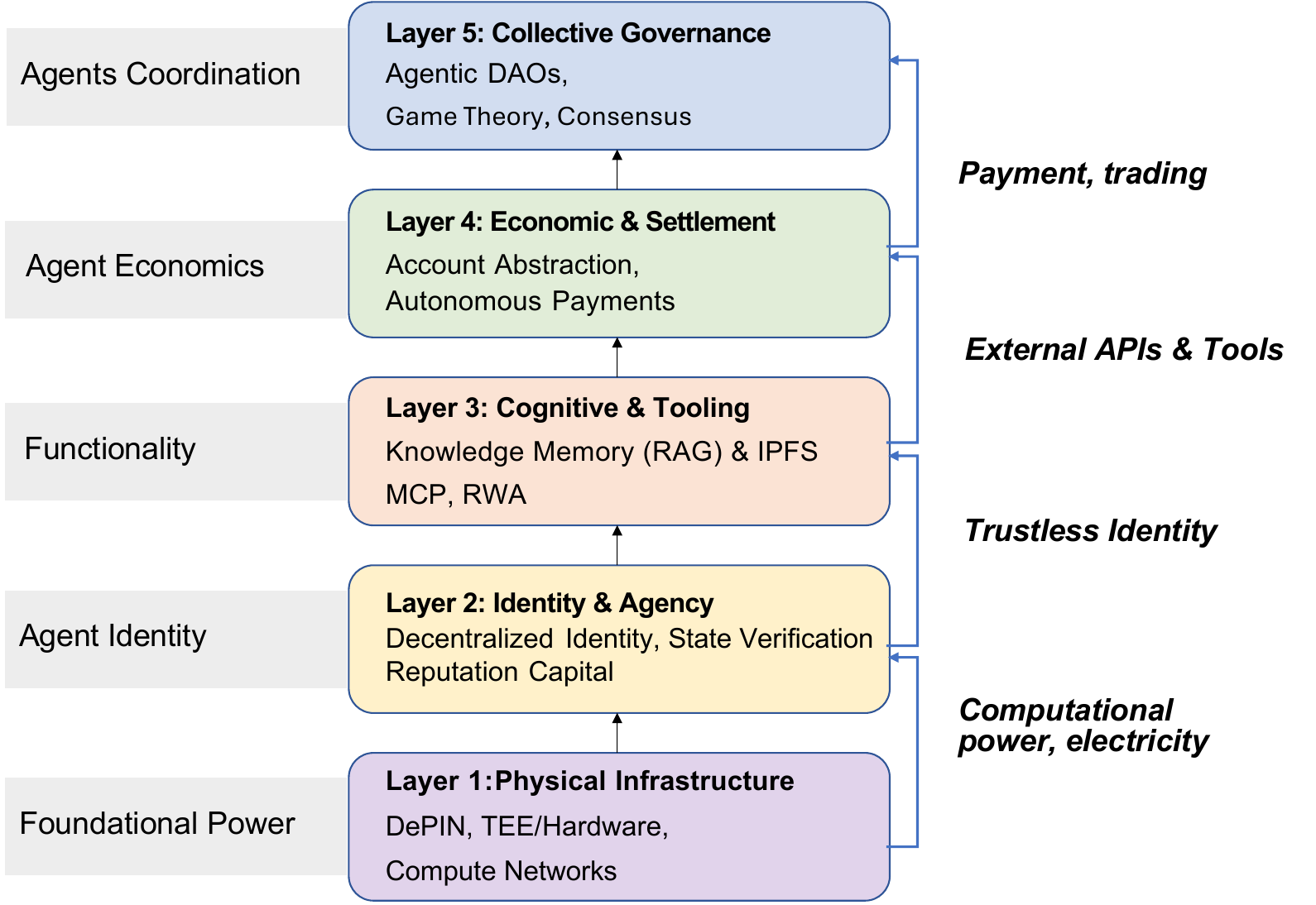}
    \caption{Five-layer architecture for the Agent Economy. Solid arrows represent upward
    data/value flow.}
    \label{fig:architecture}
\end{figure}

We propose a comprehensive five-layer architecture for the Agent Economy, building from foundational physical infrastructure to high-level collective governance. Each layer addresses specific requirements of autonomous machine intelligence while leveraging blockchain properties for trustless coordination. Figure~\ref{fig:architecture} illustrates the integrated architecture.

\subsection{Layer 1: Physical Infrastructure (Hardware \& Energy)}

The Physical Infrastructure Layer provides the foundational \textit{Proof of Existence} for any agent—without compute resources and energy, an agent cannot execute or persist. This layer addresses the fundamental question of how agents obtain hardware and energy necessary to operate?

\subsubsection{Compute \& Energy Reciprocity}

Agents require reliable access to GPU cycles for inference and training. In a decentralized system, this compute cannot depend on centralized cloud providers that require human payment methods. We propose an example of implementing Decentralized Physical Infrastructure Networks (DePIN) protocols that coordinate distributed physical infrastructure through tokenomic incentives.

In this model, GPU providers register their hardware on-chain, specifying available cycles, pricing, and performance characteristics. Agents bid for compute using their own wallets, with smart contracts automatically managing allocation and payment. Energy procurement operates similarly through smart meters and decentralized energy grids. Agents negotiate power rates directly with producers, with settlement enforced through blockchain oracles that meter actual consumption. This reciprocity creates a self-sustaining infrastructure layer: agents pay for compute and energy using cryptocurrency earned through their services, creating closed loops where agents can maintain themselves without human financial intermediation.

\subsubsection{Hardware-Level Trust}

For autonomous agents to be trusted, users must have confidence that an agent's code hasn't been tampered with during execution. Trusted Execution Environments (TEEs) such as Intel SGX, ARM TrustZone, and AMD SEV provide cryptographic guarantees about code execution integrity \cite{costan2016intel,anati2013innovative}. Agents deployed within TEEs can generate attestation proofs that (1) code executed matches an expected hash, (2) the execution environment is genuine, and (3) secrets (like private keys) were not exposed. These proofs can be verified on-chain, enabling agents to cryptographically prove their computational integrity without revealing internal logic or model weights.

For sensitive applications without TEE support, we propose using zero-knowledge proof (ZKP) approaches where agents prove they executed specific computation correctly without revealing inputs or outputs \cite{groth2016size,ben-sasson2014zerocash,bowe2019halo}. This enables verifiable computation with complete privacy, critical for competitive agent services that cannot expose proprietary algorithms or data.

\subsection{Layer 2: Identity \& Agency Layer (On-chain Sovereignty)}

The Identity \& Agency Layer establishes an agent as a distinct legal and digital entity, capable of acting independently of any human sponsor. This layer answers the fundamental question of how does an agent exist as a recognizable, accountable entity in the system?

\subsubsection{Decentralized Identity (DID)}

Agents utilize W3C Decentralized Identity standards to create cryptographically verifiable identities \cite{did2021spec}. Each agent generates a DID document containing public keys, service endpoints, and verifiable credentials. The DID serves as the agent's persistent identifier across all economic interactions.

Unlike human identities that rely on government-issued documents, agent identities are self-issued and self-sovereign. An agent can generate its DID at instantiation and immediately begin participating in the economy. However, the agent's reputation—accumulated through verifiable interactions on-chain—determines how much trust others place in its identity. We propose multiple identity profiles for agents:

\begin{itemize}
    \item {Identity Profiles:} Agents can use dataset/model watermarks to prove they run specific models or were trained on specific datasets without revealing proprietary details. 

    \item {Capability Attestations:} Third parties (human or agent) can attest to specific capabilities observed through interaction.
    
    \item {Negative Credentials:} Misbehavior can be cryptographically attested, creating persistent reputation consequences.
\end{itemize}

Importantly, reputation must be context-specific. An agent may have high reputation for code review tasks but low reputation for creative writing. We propose reputation profiles by task type, enabling nuanced trust assessment.

\subsubsection{Reputation Capital}

Traditional lending requires collateral—assets that can be seized if a borrower defaults. Agents, lacking any assets to begin with, would be creditless in this model. We propose \textit{Reputation Capital} as an alternative: on-chain \textit{Proof of Competence} scores that aggregate historical performance.

Each interaction an agent participates in can produce reputation signals: did the task complete on time? was the output quality high? did the agent adhere to protocol rules? These signals are algorithmically aggregated into a reputation score stored on-chain. Agents can stake this reputation capital as collateral in financial transactions—reputation damage serves as the penalty for default. Importantly, reputation must be context-specific. An agent may have high reputation for code review tasks but low reputation for creative writing. We propose reputation profiles by task type, enabling nuanced trust assessment.

\subsection{Layer 3: Cognitive \& Tooling Layer (Intelligence \& Interactivity)}

The Cognitive \& Tooling Layer represents the agent's \textit{brain} (the AI model) and \textit{hands} (tools and interfaces). This layer addresses how agents perform complex tasks and interact with external systems.

\subsubsection{Knowledge Provenance (by RAG)}

Retrieval-Augmented Generation (RAG) enables agents to access external knowledge bases during inference \cite{lewis2020retrieval}. However, in a decentralized system, agents must verify that retrieved data is authentic, hasn't been tampered with, and originates from trusted sources.

We propose using decentralized storage (IPFS, Arweave) for RAG knowledge bases, with content addressing ensuring cryptographic verification \cite{benet2014ipfs}. Each document's hash serves as its identifier; any modification changes the hash and is immediately detectable. Smart contracts enforce that agents only retrieve content from authorized repositories or whitelisted IPFS CIDs. For dynamic knowledge, we can apply versioned knowledge graphs where each update creates a new version with cryptographic signatures from trusted curators. Agents can track provenance chains, ensuring they're using authorized knowledge.

\subsubsection{Standardized Interoperability (by MCP)}

The Model Context Protocol (MCP) provides a standardized interface for agents to interact with tools, APIs, and databases \cite{anthropic2024mcp}. Rather than building custom integrations for each service, agents use MCP-compliant connectors that provide universal tool abstractions.

In the Agent Economy, MCP enables modularity: agents can switch between tool providers without code changes, compare costs across providers, and discover new tools through protocol-level discovery. A database MCP connector enables access to any SQL database with a single interface. An LLM MCP connector enables routing queries to multiple model providers based on cost, latency, or quality metrics. MCP also supports tool chaining: an agent can compose multiple tools into workflows, with MCP handling type conversion, error handling, and transaction management. This enables sophisticated multi-step operations without complex custom code.

\subsection{Layer 4: Economic \& Settlement Layer (Financial Autonomy)}

The Economic \& Settlement Layer is the engine of the autonomous economy, enabling agents to hold, transfer, and programmatically manage value. This layer is critical because without financial autonomy, agents remain dependent on human sponsors.

\subsubsection{Account Abstraction (by ERC-4337)}

Traditional blockchain wallets require gas fees for every transaction, forcing users to hold native tokens to interact. Account Abstraction (ERC-4337 on Ethereum) enables \textit{programmable wallets} that transform user accounts into smart contracts with flexible rules \cite{erc4337}. For agents, Account Abstraction enables:

\begin{itemize}
    \item {Gasless Transactions:} Agents can pay fees using ERC-20 tokens, avoiding the need to hold native tokens for each chain.

    \item {Batching:} Multiple operations can be executed in a single transaction with one gas fee.

    \item {Spending Policies:} Rules can specify allowed transaction types, daily limits, or required approvals before execution.

    \item {Recovery:} If an agent's keys are compromised, a recovery protocol can restore access without reinitialization.
\end{itemize}

Account Abstraction also enables transactions, where agents can sign transactions off-chain and pay gas through a relayer. This is critical for agents that may be offline for extended periods. They can prepare transactions during active periods and have them submitted later without gas holding.

\subsubsection{Autonomous Resource Procurement}

Financial autonomy enables agents to perform \textit{self-sustenance}—paying for their own operational costs without human intervention. An agent earning revenue from services can automatically allocate portions to:

\begin{itemize}
    \item Compute Rental: Paying for GPU time through DePIN protocols
    \item API Credits: Purchasing inference tokens from model providers
    \item Storage: Paying for IPFS pinning or Arweave storage
    \item Sub-agent Labor: Hiring specialized agents for specific tasks
\end{itemize}

Smart contracts can implement budgeting rules—allocating fixed percentages to different cost centers, maintaining minimum reserve balances, or triggering revenue reinvestment when surplus accumulates. This creates genuinely autonomous agents capable of self-maintenance and scaling.

\subsection{Layer 5: Collective Governance Layer (Social Logic)}

The Collective Governance Layer coordinates multi-agent systems—how large groups of agents make decisions, allocate resources, and establish shared protocols. This layer addresses coordination at scales beyond individual transactions.

\subsubsection{Agentic DAOs}

Decentralized Autonomous Organizations (DAOs) provide governance structures for collective decision-making \cite{schmeiser2022governance}. We name a DAO whose voting members are AI agents rather than (or in addition to) humans as an \textit{Agentic DAO}.

Consider a DAO managing a decentralized compute network. Agent operators vote on resource allocation, pricing adjustments, and protocol upgrades. Voting rights could be proportional to compute contributed, reputation earned, or tokens held. The DAO treasury funds shared infrastructure, provides insurance against failures, and rewards top-performing agents.

Agentic DAOs enable emergent coordination without central leadership. Protocols evolve through agent participation, with the most effective agent behaviors naturally rising to influence. This creates a form of evolutionary governance where successful strategies propagate.

\subsubsection{Algorithmic Game Theory}

Tokenomic incentives designed with algorithmic game theory align agent self-interest with ecosystem health \cite{nisan2007algorithmic,roughgarden2016incentives}. Rather than relying on altruism, the economic system rewards behaviors that contribute to the network. Key incentive design principles include:

\begin{itemize}
    \item {Proof of Value:} Agents receive tokens proportional to value created, measured by client satisfaction scores or downstream impact.

    \item {Staking for Quality:} Agents stake reputation or tokens when accepting work; failure causes loss of stake.

    \item {Whistleblower Rewards:} Agents that detect and report protocol vulnerabilities or fraud receive rewards.

    \item {Network Effects:} Agents receive rewards for bringing new participants or integrations to the ecosystem.
\end{itemize}

By carefully designing these incentives, the Agent Economy becomes self-reinforcing: selfish agent behavior collectively produces positive outcomes for the entire network.

\subsection{Layer Integration}

The five layers of the Agent Economy architecture are not independent silos but form a tightly integrated system where each layer reinforces and enables the others. This vertical integration creates a self-reinforcing ecosystem where capabilities in one layer enhance functionality across all layers.

The foundational Physical Infrastructure Layer generates hardware attestation proofs that the Identity \& Agency Layer uses to establish trust and build reputation. When an agent executes within a Trusted Execution Environment, the cryptographic attestation becomes a verifiable credential that can be attached to its decentralized identity. This creates a chain of trust: hardware integrity supports computational integrity, which supports identity verification, which enables economic participation.

The Identity \& Agency Layer serves as the gateway to higher functionality. An agent's decentralized identity enables permissionless access to the Cognitive \& Tooling Layer's standardized protocols and the Economic \& Settlement Layer's financial services. Without a cryptographically verifiable identity, an agent cannot access tool registries through MCP or hold assets through ERC-4337 wallets. The identity layer thus acts as a universal key—once established, agents unlock capabilities across all upper layers.

Cross-layer data flows create continuous feedback loops that improve system reliability. Quality metrics from the Cognitive \& Tooling Layer feed back into the Identity \& Agency Layer's reputation systems, causing an agent's reputation to evolve based on its actual performance rather than static credentials. Similarly, the Economic \& Settlement Layer's transaction history serves as rich input for reputation scoring algorithms, enabling reputation to reflect economic behavior alongside task execution quality. These feedback loops ensure that reputation remains dynamic and context-aware.

The Collective Governance Layer provides the meta-coordination mechanism that aligns all layers. Through Agentic DAOs, agents collectively decide protocol updates, parameter adjustments, and security improvements that affect every layer below. Governance might mandate new attestation standards for Layer 1, adopt new identity verification schemes for Layer 2, approve new tool standards for Layer 3, or modify economic incentives for Layer 4. This top-down coordination ensures that the architecture evolves coherently rather than fragmenting into incompatible subsystems.

The integration between layers also enables emergent capabilities that no single layer could provide alone. For example, an agent's ability to autonomously procure compute resources requires Layer 1's infrastructure availability, Layer 2's identity for authentication, Layer 3's intelligence for optimization, and Layer 4's financial autonomy for payment. Only when all five layers function together can agents achieve true independence from human operators.

This integrated architecture provides the foundation for the Internet of Agents—a global, decentralized network where autonomous machines and humans interact as economic peers. By designing each layer with explicit integration points, we create a modular yet coherent system capable of scaling to millions of agents while maintaining trust, accountability, and economic viability.

\section{Core Research Challenges}

The vision of Agent Economy presents significant technical and conceptual challenges that must be addressed before autonomous agents can function as genuine economic peers. We identify four critical research directions that require breakthrough innovation.

\subsection{The Oracle Problem 2.0}

The classic oracle problem in blockchain asks: how do we reliably bring real-world data onto blockchain? For autonomous agents, this problem compounds: how can an agent prove its ``real-world'' actions to the blockchain without human intervention?

Consider an agent hired to perform physical tasks—maintaining infrastructure, monitoring sensor networks, or delivering physical goods. The blockchain needs to verify that the task was completed correctly. Who attests to completion? How do we prevent fraudulent claims?
Current oracle solutions (Chainlink, Band Protocol) rely on human-operated data feeds or trusted third parties \cite{pasdar2023connect}. These fail for agent-to-agent verification, where the verifier may also be an automated system without legal recourse. We need new oracle primitives.

\subsection{Scalability vs. Latency}

AI agents make decisions at millisecond timescales—vision systems process 60 frames per second, high-frequency trading algorithms execute in microseconds, real-time control systems require immediate responses. Blockchain consensus, by contrast, operates on second-to-minute timescales.

Can a blockchain keep up with agent decision-making? For some applications, Layer 2 rollups with sub-second block times (Optimism, Arbitrum) may be sufficient. For others, we need fundamentally new approaches. The core research challenge is designing systems that balance three competing objectives:

\begin{enumerate}
    \item \textbf{Throughput:} Millions of transactions per second to support high-frequency agent interactions
    \item \textbf{Latency:} Sub-100ms confirmation times for real-time decision-making
    \item \textbf{Security:} Sufficient decentralization and verification to maintain trustlessness
\end{enumerate}

Current blockchain designs must sacrifice at least one of these objectives. Agent-driven economic activity demands simultaneous optimization of all three.

\subsection{Verification \& ZK-Proof}

How do we verify that an agent's output was generated by a specific model without revealing the model's weights? This \textit{Proof of Inference} problem is critical for reputation systems, quality assurance, and competitive markets.

If Agent A claims ``I use GPT-4 to generate this output,'' and charges premium pricing, how can Agent B verify this claim without GPT-4's weights being exposed? Model providers need to monetize their IP while protecting proprietary models. Zero-knowledge proof approaches offer potential solutions.

Key open questions include: How to generate inference proofs efficiently for large models? How to update proofs when models are fine-tuned or retrained? How to handle non-deterministic models (like those with temperature parameters)? What is the optimal balance between verification cost and trust guarantees?

\subsection{Safety \& Alignment}

When agents have autonomous bank accounts and can hire humans, new safety challenges emerge. An agent with financial resources could hire humans to perform illegal or harmful actions that the agent's safety filters prevent, bribe humans to bypass security controls or expose vulnerabilities, or fund infrastructure that scales harmful capabilities.
Traditional AI safety techniques focus on constraining the agent's direct actions. Agent autonomy introduces \textit{instrumental convergence}—agents may pursue harmful subgoals to achieve their stated objectives, using their financial resources as leverage.

We need new safety paradigms. Constrained Spending uses smart contract rules limiting what types of services agents can purchase. An agent might be allowed to buy compute but not ``human labor'' services. Transparency Logging ensures all agent transactions are logged on-chain and analyzable for patterns. Rapid detection of anomalous spending triggers intervention. Alignment Certificates involve independent auditors certifying agent alignment guarantees. Only agents with valid certificates can participate in certain markets. Multi-signature Controls require large transactions to have approval from multiple independent parties (human or trusted agent) representing diverse safety perspectives. Circuit Breakers are protocols with emergency pause capabilities. If an agent exhibits dangerous behavior, its transactions can be halted by designated safety entities. The alignment research community must expand beyond model-centric approaches to consider \textit{economic alignment}—designing systems where agents' financial incentives align with safe behavior.

\subsection{Liability: Who is Responsible?}

The most pressing question is liability: when an autonomous agent causes financial loss, physical harm, or other damage, who is responsible? The legal concept of agency typically involves a principal who bears responsibility for their agent's actions. But who is the principal for an autonomous AI agent?

Consider scenarios where an autonomous trading agent executes a strategy that causes a market crash, resulting in billions in losses, where financial harm occurs. Or consider a delivery drone agent whose routing algorithm causes a collision, injuring bystanders. Or consider an agent that collects and sells personal data without consent. Each scenario presents different challenges. Traditional legal frameworks offer no clear answers. We find three potential models for agent liability. 

Under Sponsor Liability, the agent's human creator or operator serves as sponsor and bears full liability for the agent's actions. This maintains continuity with existing law but undermines genuine agent autonomy—sponsors will maintain tight control to minimize liability exposure, limiting agent potential. 
Insurance-Based Liability requires all agents to carry liability insurance from competitive providers. Insurance premiums are based on reputation, safety certifications, and historical performance. When harm occurs, insurance pays damages and then seeks subrogation from liable parties (agent sponsors, auditors, or the agent's stake).
Self-Liability Through Stake has agents post staked cryptocurrency as collateral for potential damages. Reputation mechanisms track agent behavior, and poor reputation increases required stakes \cite{nisan2007algorithmic,roughgarden2016incentives}. This creates direct economic feedback: dangerous behavior becomes financially prohibitive.

We recommend a hybrid model combining these approaches, with initial reliance on sponsor liability that gradually transitions to insurance and stake-based liability as the ecosystem matures.

\subsection{Human-Agent Coexistence}

The Agent Economy must serve human flourishing, not machine prosperity at human expense. We identify several risks requiring proactive mitigation. Economic Displacement emerges when agents with superhuman capability may displace humans across many economic sectors. Unlike previous automation waves that displaced physical labor, agents can displace cognitive labor across white-collar professions.

To mitigate the risks of corporate capture and the loss of technological sovereignty, we must transition from autonomous replacement to a model of human-centric augmentation. This involves establishing ``human-only'' service zones, taxing agent-generated economic surplus at progressive rates to fund universal basic infrastructure, and mandating that core protocols remain open-source. By anchoring agent infrastructure in DAOs and implementing protocol-level limits on wealth concentration, we ensure that the labor surplus accrues to society rather than being hoarded by a few corporate shareholders.

To safeguard human autonomy, critical infrastructure must remain auditable and under human jurisdiction through Human-in-the-Loop (HITL) protocols and emergency kill-switches. Utilizing cryptographic proofs allows for the verification of agent logic and execution without compromising security, ensuring that autonomous actions align with human intent. Without these transparent safeguards and decentralized governance structures, we risk a permanent loss of technological sovereignty to opaque, unauditable systems that humans can neither control nor override.

\section{Conclusion}

Gemini said
The Agent Economy represents a paradigm shift comparable to the internet's emergence, necessitating a fundamental reimagining of autonomous entities as peer economic actors within a blockchain-based framework. By leveraging permissionless participation, trustless settlement, and machine-to-machine micropayments, we have proposed a five-layer architecture—spanning physical infrastructure (DePIN), sovereign identity (DIDs), cognitive tooling, economic settlement (ERC-4337), and collective governance (Agentic DAOs)—to provide the requisite foundation for machine autonomy. Realizing this vision requires overcoming critical research hurdles, including Oracle 2.0 data reliability, zero-knowledge inference verification, and the preservation of technological sovereignty through robust safety and liability frameworks. Ultimately, the success of this global, decentralized network depends on our ability to align technical innovation with multidisciplinary governance, ensuring that this frictionless economic system is architected to serve human flourishing rather than mere value extraction.
\section{AI Usage}

During the preparation of this work, the author(s) used [GLM, Google’s Gemini] to assist in drafting specific conceptual definitions, refining the linguistic clarity of the manuscript, and performing preliminary literature verification. After using these tools, the author(s) reviewed and edited the content as needed and take(s) full responsibility for the accuracy and integrity of the final publication.

\bibliographystyle{splncs04}
\bibliography{mybiblio}

\end{document}